\documentclass[aps,twocolumn,showpacs,preprintnumbers,nofootinbib,prd,superscriptaddress,10pt]{revtex4-2}
\makeatletter
\def\l@subsubsection#1#2{}
\def\l@subsubsubsection#1#2{}
\makeatother

\usepackage{graphicx,amssymb,amsmath,amsthm,amsfonts,epsfig,epsf,fixmath}
\usepackage[usenames]{color}
\usepackage{epstopdf}

\usepackage{aas_macros}
\usepackage{dcolumn}
\usepackage{latexsym}
\usepackage{rotating}
\usepackage{longtable}

\usepackage{enumerate}
\usepackage{url}
\usepackage[linktocpage]{hyperref}
\usepackage{isotope}

\newcommand{\tn}{\textnormal}

\begin{document}
\title{Constraining three-nucleon forces with multimessenger data}
\author{Andrea Maselli}
\affiliation{Gran Sasso Science Institute (GSSI), I-67100 L’Aquila, Italy}
\affiliation{INFN, Laboratori Nazionali del Gran Sasso, I-67100 Assergi, Italy}
\affiliation{Dipartimento di Fisica, ``Sapienza'' Universit\`a di Roma, Piazzale
Aldo Moro 5, 00185, Roma, Italy}
\author{Andrea Sabatucci}
\affiliation{Dipartimento di Fisica, ``Sapienza'' Universit\`a di Roma, Piazzale
Aldo Moro 5, 00185, Roma, Italy}
\affiliation{INFN, Sezione di Roma, Piazzale
Aldo Moro 5, 00185, Roma, Italy}
\author{Omar Benhar}
\affiliation{INFN, Sezione di Roma, Piazzale
Aldo Moro 5, 00185, Roma, Italy}
\affiliation{Dipartimento di Fisica, ``Sapienza'' Universit\`a di Roma, Piazzale
Aldo Moro 5, 00185, Roma, Italy}

\date{\today}

\begin{abstract}
We report the results of a study aimed at inferring direct 
information  on the repulsive three-nucleon potential 
$V^R_{ijk}$\textemdash driving the stiffness of the nuclear 
matter equation of state at supranuclear densities\textemdash 
from astrophysical observations. Using a Bayesian approach, 
we exploit the measurements of masses, radii and tidal 
deformabalities performed  by the NICER satellite and the 
LIGO/Virgo collaboration, as well as the mass of the heaviest 
observed pulsar, to constrain the strength of $V^R_{ijk}$. The 
baseline of our analysis is the widely employed nuclear Hamiltonian 
comprising the Argonne $v_{18}$ nucleon-nucleon potential and
the Urbana IX model of three-nucleon potential. 
The numerical results, largely determined by the bound on the maximum mass, 
suggest that existing and future facilities have the potential 
to provide valuable new insight into microscopic nuclear 
dynamics at supranuclear densities.
\end{abstract}

\maketitle

\section{Introduction}
Over the past decade, the availability of astrophysical 
data collected by electromagnetic (EM) observatories 
\cite{Demorest:2010bx,Antoniadis:2013pzd,Fonseca:2016tux,Cromartie:2019kug,
Ozel:2010fw,Steiner:2010fz,Guver:2013xa,Riley:2019yda,Miller:2019cac} 
and gravitational-wave (GW) interferometers 
\cite{Abbott:2018wiz,Abbott:2020uma,TheLIGOScientific:2017qsa},  
supplemented by the information obtained from Earth-based 
laboratory experiments \cite{symmetry:expt,russotto,Tsang,IAS,Brown,Zhang,danielewicz,Shlomo2006,Colo2008,Shlomo2006}, 
has opened a new era for the investigation of neutron star (NS) 
structure and dynamics. 

The studies aimed at constraining the equation of state (EOS) of NS 
matter have recently benefit from measurements 
of the tidal deformabilities \cite{Hinderer:2007mb,Damour:2009vw,Binnington:2009bb,Flanagan:2007ix,Vines:2010ca,Vines:2011ud}\textemdash
encoding the footprint of tidal interactions on the 
signal emitted by a binary system\textemdash performed within the GW band. 
Because the tidal deformability 
depends on the internal composition of the stars, 
any information on its value is potentially a source of novel 
insight into the EOS. The discovery of GW170817 has triggered 
a large number of efforts aimed at constraining the NS 
structure, also exploiting multimessenger approaches 
based on joint GW-EM analyses \cite{Annala:2017llu,Margalit:2017dij,Radice:2017lry,Bauswein:2017vtn,Lim:2018bkq,Most:2018hfd,Carson:2018xri,De:2018uhw,Annala:2019puf,Raaijmakers:2019dks,Miller:2019nzo,Kumar:2019xgp,Fasano:2019zwm,Guven:2020dok,Traversi:2020aaa,Landry:2020vaw,Zimmerman:2020eho,Silva:2020acr, Sabatucci2020}. 
For extensive reviews on this topic, see Refs.~\cite{Baiotti2019,Chatziioannou:2020pqz} and references therein.

Besides being a valuable source of information on average 
properties of dense matter embodyed in the EOS,  
as well as on the possible occurrence of a transition 
to exotic high-density phases\textemdash in which constituents 
other than nucleons are the relevant degrees of 
freedom\textemdash the new data provide 
an unprecedented opportunity to constrain the existing models 
of nuclear dynamics at supranuclear density.

The description of nuclear systems in terms of point-like 
protons and neutrons interacting through phenomenological
two- and three-body forces\textemdash hereafter 
Nuclear Many-Body Theory, or NMBT\textemdash has proved 
remarkably successful. The results of calculations based on NMBT 
account for a variety of observables 
of nuclei with ${\rm A \leq 12}$\textemdash including the 
energies of ground and low-lying excited states and 
electromagnetic form factors~\cite{QMC}\textemdash 
and provide accurate and reliable estimates of the empirical equilibrium 
properties of isospin-symmetric matter~\cite{APR}.
Applications of NMBT in the region of supranuclear densities, however,
unavoidably involve a degree of extrapolation.

Phenomenological NN potentials account for scattering 
data up to  large energies\textemdash typically  $\sim$~600 MeV 
in the lab frame, well beyond pion production treshold\textemdash as well as for the effects of high-momentum components 
in the deuteron wave-function. Therefore, these models are
expected to be suitable to describe interactions in matter at densities 
as large as $\sim 4 \varrho_0$, with $\varrho_0$ being the nuclear 
matter equilibrium density~\cite{Benhar_EFT}. 

Three-nucleon (NNN) potentials, on the other hand, are designed 
to explain only the ground-state energy of the three-nucleon 
bound states and nuclear matter equilibrium properties, while being
totally unconstrained in the regime 
corresponding to densities $\varrho \gg \varrho_0$, in which 
the contribution of three-nucleon interactions rapidly increases to 
becomes dominant. 
In this article, we report the results of an analysis based on a novel approach, 
in which astrophysical data are exploited to obtain information 
on the strength of the repulsive component of the NNN potential, 
which largely determines the compressibility of matter at supranuclear densities. 

It has to be stressed that our approach is conceptually different
from the one aimed at constraining  the EOS of dense matter, and
has a much broader scope. The bottom line is that, while being 
admittedly non observable, the NN and NNN potentials 
can be modelled in such a way as to reproduce the available data, 
and the resulting phenomenological Hamiltonians allow to perform 
calculations ofa variety of nuclear matter properties other than 
the ground-state energy, whose understanding is needed
for a comprehensive  description of neutron star structure and dynamics. 
This line has been pursued by the authors of Refs.~\cite{BV,BPVV,BF,LBGL,BDR},  
who carried out pioneering studies of the transport coefficients, the response 
to neutrino interactions, and the superfluid gap of nuclear mater 
based on a unified microscopic description of nuclear interactions.

The body of this article is structured as follows. In 
Sect.~\ref{dynamics} we outline the dynamical model underlying our 
study, as well as the simple parametrisation adopted to 
characterise the strength of the repulsive component of the NNN 
potential. The datasets considered in the analysis are described in Section~\ref{data}, while the results are reported and discussed 
in Sect.~\ref{results}. Finally, the summary of our findings and 
an outlook to future developments can be found in in Sect.~\ref{summary}.
Unless explicitly stated otherwise, we shall use geometric ($G=c=1$) 
units.

\section{Dynamical model}
\label{dynamics}
The model of nuclear dynamics employed in our work is based on a Hamiltonian of the form
\begin{equation}
H = \sum_{i}\frac{p_i^2}{2m} + \sum_{i<j}v_{ij}+\sum_{i<j<k}V_{ijk} \ . 
\end{equation} 
Here, $v_{ij}$ is the Argonne $v_{18}$ (AV18) nucleon-nucleon 
(NN) potential~\cite{AV18}, corrected to take into account 
relativistic boost corrections needed to describe NN interactions 
in the locally inertial frame associated with a star~\cite{Forest,APR}. 
The NNN potential, $V_{ijk}$, is assumed to consist of a two-pion 
exchange contribution, $V^{2 \pi}_{ijk}$, and a purely phenomenological 
repulsive term, $V^{R}_{ijk}$ which largely determines the stiffness 
of the nuclear matter EOS at high densities, usually parametrised 
by the compression  modulus $K$. In the Urbana IX (UIX) 
model~\cite{Urbana_model}, the strengths of $V^{2 \pi}_{ijk}$ 
and $V^{R}_{ijk}$ are fixed in such as way as to reproduce the 
ground-state energy of $\isotope[3][]{\rm H}$ and the saturation 
density of nuclear matter, respectively. The Hamiltonian comprising 
the boost-corrected AV18 potential and the UIX potential has been 
employed to obtain the EOS of Akmal, Pandharipande and Ravenhall 
(APR)~\cite{APR}, widely used in calculations of neutron star 
properties. The pressure of isospin-symmetric nuclear 
matter computed using this Hamiltonian turns out to be 
consistentwith the bounds obtained from the analysis of
high-energy nuclear collisions of Ref~\cite{danielewicz}, 
extending to  $\varrho \gtrsim 4 \varrho_0$.

In order to explore the possibility to constrain the NNN 
potential using astrophysical data, we have generated a 
set of EOSs, computed replacing  $V^{R}_{ijk} \to \alpha V^{R}_{ijk}$ 
and varying the value of the parameter $\alpha$, which 
determines the strength of the repulsive NNN interaction. 
For any values of $\alpha$, the EOS of charge neutral and 
$\beta$-stable matter consisting of protons, neutrons, 
electrons and muons has been obtained using a straightforward 
generalisation of the fitting procedure discussed in 
Ref.~\cite{APR} yielding the energy-density functional 
${\mathcal H}_\alpha(\varrho,x)$, with $x$ being the proton 
fraction. Note that, while the EOS of Ref.~\cite{APR} has 
been obtained within a variational approach, in our work 
the contributions arising from $(1-\alpha) V^{R}_{ijk}$ 
are added in perturbation theory. To maintain consistency 
with the APR EOS, corresponding to $\alpha = 1$, at 
densities $\varrho \approx \varrho_0$, we have limited 
our analysis to the range $0.7 \leq \alpha \leq 2$. Within 
these bounds the displacement of the equilibrium density 
with respect to the empirical value    
remains small, ranging between $\sim 4 \%$ 
and $\sim 20 \%$. 
This range appears to be acceptable, 
considering that the corresponding change of the energy per 
particle does not exceed $3 \%$. Moreover, the pressure of 
isospin-symmetric matter turns out to be compatible with the 
empirical constraints derived in Ref.~\cite{danielewicz} for 
all values of $\alpha$. 

\begin{figure}[!htbp]
\centering
\includegraphics[width=0.55\textwidth]{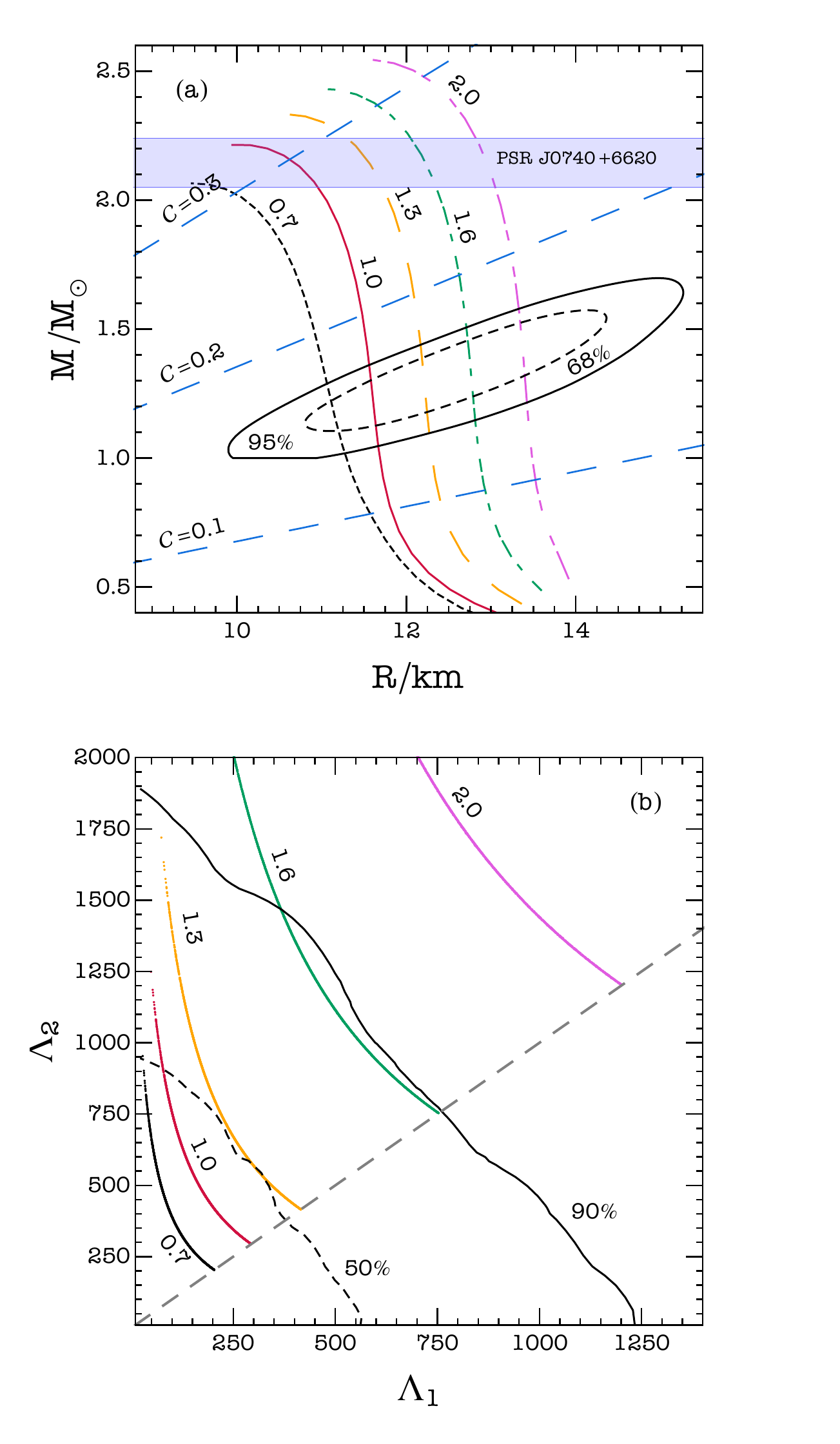}
\caption{(a) Mass-radius relations of NS  
obtained using EOSs corresponding to the values of  
$\alpha$ specified on top of each curve. The shaded 
band identifies the most massive pulsar observed 
so far, PSR J0740+6620 \cite{Cromartie:2019kug}.
The closed contours show to the 68\% and 95\% 
confidence intervals derived for the NICER pulsar 
\cite{Riley:2019yda}. The dashed straight lines correspond to 
constant compactness ${\cal C}=M/R$. (b) NS models 
in the $\Lambda_1$-$\Lambda_2$ plane for selected values 
of $\alpha$, and corresponding to the masses measured 
for GW170817. The 50\% and 90\% confidence 
intervals derived using the GW170817 data 
\cite{TheLIGOScientific:2017qsa} are also shown. \label{fig:MR}}
\end{figure}

\section{Astrophysical datasets}
\label{data}
The EOSs belonging to the family considered in our work only 
depend on the strength of the NNN force. Therefore, stellar 
configurations are specified by the values of $\alpha$ and 
the central pressure $p_c$, and provide two macroscopic 
observables: the mass $M(\alpha,p_c)$ and either the radius, 
$R(\alpha,p_c)$, or the tidal deformability\footnote{We refer
here to the \textit{normalised} tidal deformability
$\Lambda=\lambda/M^5$, where $\lambda=2/3k_2 R^5$
is the quadrupolar Love number.}, $\Lambda(\alpha,p_c)$, 
to be compared with observations.

We consider two classes of datasets: (i) the GW observation 
of the binary NS event\footnote{The LIGO/Virgo collaboration 
has recently announced the detection of GW190814, an 
asymmetric binary system featuring a black hole and a $2.6 \ M_\odot$
companion \cite{Abbott:2020khf}. While it is sitll unclear
whether the latter is a NS or a black hole, this discovery
could have a profound impact on our understanding of stellar
evolution, see, e.g., Refs.~\cite{Most:2020bba,Fattoyev:2020cws,Clesse:2020ghq,Tews:2020ylw,Lim:2020zvx,Biswas:2020xna}.} GW170817 
\cite{TheLIGOScientific:2017qsa}, and (ii) the
spectroscopic observation in the EM bandwidth of the 
millisecond pulsars PSR J0030+0451 performed by the NICER 
satellite \cite{Riley:2019yda}. Figure~\ref{fig:MR} shows 
the confidence intervals in the $M$-$R$ (a) and $\Lambda_{1}$-$\Lambda_2$ (b)
plane for the two datasets considered, together with the 
results corresponding to stellar configurations having 
different values of $\alpha$. It appears that, compared to 
NICER results, the tidal deformabilities inferred from GW170817 
tend to rule out models with $\alpha\gtrsim 1.6$, corresponding 
to the stiffest EOSs.

To constrain the parameters associated with the EOS, we sample 
their probability distribution through a Bayesian approach 
based on Monte Carlo Markov Chain (MCMC) simulations 
\cite{gilks1995markov}. For a given astrophysical dataset 
$O$ comprising $n$ observed stars, the probability distribution 
of $\theta=\{\alpha,p_c^{(i=1,\ldots n)}\}$, is given by
\begin{equation}
{\cal P}(\theta\vert O)\propto {\cal P}_0(\theta){\cal L}(O\vert D(\theta))\ ,\label{math:bayes}
\end{equation}
where ${\cal P}_0$ is the prior information on $\theta$,
and ${\cal L}$ is the likelihood function, with $D(\theta)$
being the set of NS observables needed to interpret the data,
i.e. the mass, the radius and/or the tidal deformability.
For the GW event the likelihood function is given by the marginalised 
joint density distribution\footnote{The likelihood function can 
be directly derived from the GW posterior since the joint prior 
on the masses and the tidal deformabilities is flat. The same 
holds for the NICER dataset, since the joint prior on the mass 
and the radius is also flat.} inferred by LIGO/Virgo 
\cite{Ind:eos}:
\begin{equation}
{\cal L}(O_\tn{GW}\vert D(\theta))={\cal L}_\tn{GW}(M_1,M_2,\Lambda_1,\Lambda_2)\ .\label{math:LVClike}
\end{equation}
Morerover, given the exquisite precision to which the chirp
mass ${\cal M}=(M_1M_2)^{3/5}/(M_1+M_2)^{1/5}$ of GW170817
has been measured, compared to the individual masses, and following
Ref.~\cite{Raaijmakers:2019dks}\textemdash see, in particular, the 
discussion of Appendix B\textemdash  we fix it to the
median value ${\cal M}=1.186 \ M_\odot$. The mass $M_2$, as well 
as the tidal deformability $\Lambda_2$, is then uniquely determined
by $M_1$, i.e. by the central density of the primary object,
and the number of parameters to be constrained is reduced.
Finally, we transform the likelihood function to the mass
ratio\footnote{Changing ${\cal L}(M_{1},M_{2},\Lambda_1,\Lambda_2) \to 
{\cal L}({\cal M},q,\Lambda_1,\Lambda_2)$  the prior 
on the chirp mass and the mass ratio, ${\cal P}_0(q,{\cal M})$, 
is no longer flat. However, as discussed in \cite{Raaijmakers:2019dks},
reweighting GW170817 so that the prior on the mass ratio
and the chirp mass is flat has a negligible effect.}
$q=M_1/M_2$, and Eq.~\eqref{math:bayes} becomes:
\begin{equation}
{\cal P}({\alpha,p_{c}^{(1)}}\vert O_\tn{GW})\propto {\cal P}_0(\alpha,p_c^{(1)},p_c^{(2)}){\cal L}_\tn{GW}(q,\Lambda_1,\Lambda_2)\ .\label{math:bayesGW}
\end{equation}
Equation~\eqref{math:bayesGW} shows that the central
pressure of the second component of the binary system 
is also needed. We reconstruct its value by inverting 
the mass relation $M_2=M_2(\alpha,p_c^{(2)})$, to lie 
within the prior support.

For the NICER dataset the likelihod is given by the 
marginalised joint posterior \cite{raaijmakers_g_2020_3711718}
${\cal L}(O_\tn{em}\vert D(\theta))={\cal L}_\tn{EM}(M,R)$, 
such that:
\begin{equation}
{\cal P}({\alpha,p_c^{(1)}}\vert O_\tn{EM})\propto {\cal P}_0(\alpha,p_c^{(1)}){\cal L}_\tn{EM}(M,R)\ . \label{math:bayesEM} 
\end{equation}

In the multi-messenger scenario, we consider the joint 
GW and electromagnetic datasets. Since the two sets of 
observations are independent, we have
\begin{align}
{\cal P}({\alpha,p_c^{(1)},p_c^{(3)}}\vert& O_\tn{GW},O_\tn{EM})\propto {\cal P}_0(\alpha,p_c^{(1)},p_c^{(2)},p_c^{(3)}) \nonumber\\
\times &{\cal L}_\tn{GW}(q,\Lambda_1,\Lambda_2){\cal L}_\tn{EM}(M,R)\ ,\label{math:bayesGWEM}
\end{align}
where, $p_c^{(1)}$ and $p_c^{(2)}$ refer to the two NSs
of GW170817, and the third pressure $p_c^{(3)}$ corresponds 
to the pulsar observed by NICER.

The values of $\alpha$ are sampled uniformly within the 
range $[0.7,2]$, while the central pressures of each star 
are drawn in the logarithm space between $\ln_{10} p_c\simeq34.58$ 
and $\ln_{10}p_c^\tn{max}(\alpha)$. The latter represents 
the maximum central pressure for a stable configuration 
of the EOS family identified by $\alpha$. The lower end 
of the prior interval for $\alpha$ is chosen such that 
the nuclear model supports NSs with masses larger than 
$1.8M_\odot$. Moreover, we ask the speed of sound 
$c_s=\sqrt{dp/d\epsilon}$ inside each stellar configuration 
to be smaller than the speed of light.

Finally, we also include in the prior information the 
maximum mass that must be supported by the EOS, provided 
by the high-precision radio pulsars timing of the binary 
PSR J0740+6620 \cite{Cromartie:2019kug}. Following the 
procedure described in \cite{Raaijmakers:2019dks,Alvarez-Castillo:2016oln,Miller:2019cac,Miller:2019nzo}, we do not impose 
an hard prior on $M_\tn{max}$. Rather, we describe the 
highest mass measurement as a normal distribution 
$N(\mu,\sigma)$ with mean $\mu=2.14 \ M_\odot$ and standard 
deviation $\sigma=0.09M_\odot$~\footnote{The actual value of 
the mass inferred by timing observations is 
$2.14^{+0.10}_{-0.09}M_\odot$. However, the asymmetry of 
the error has a negligible effect in our analys}. This practically amounts 
to adding to the MCMC parameter vector an extra central 
pressure\textemdash which we label $p^{(M)}_c$ and 
corresponds to PSR J0740+6620\textemdash and multiplying
Eqns.~\eqref{math:bayesGW}, \eqref{math:bayesEM} and
\eqref{math:bayesGWEM} by the likelihood function
${\cal L}_{J0740+6620}=N(\mu,\sigma)$. For the multimessenger
analysis this corresponds to multiply the posterior 
\eqref{math:bayesGWEM} by $N(\mu,\sigma)$

We sample the posterior distribution using the \textit{emcee}
alghoritm with stretch move \cite{Foreman_Mackey_2013}. 
For each set of data, we run one hundred walkers of $10^6$ 
samples with a thinning factor of 0.02.

\section{Results and discussion}
\label{results}
\begin{figure}[!htbp]
\centering
\includegraphics[width=0.5\textwidth]{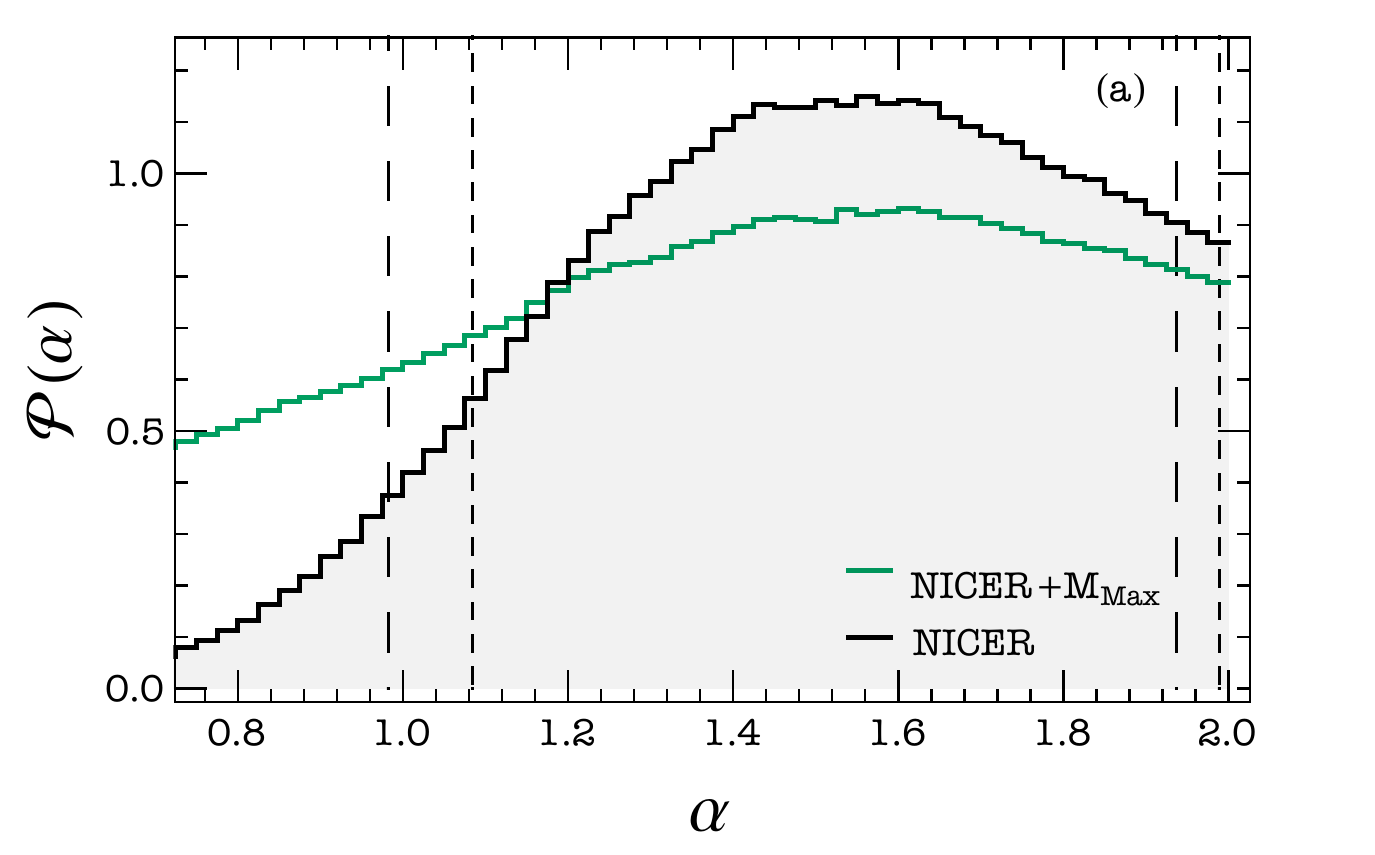}
\includegraphics[width=0.5\textwidth]{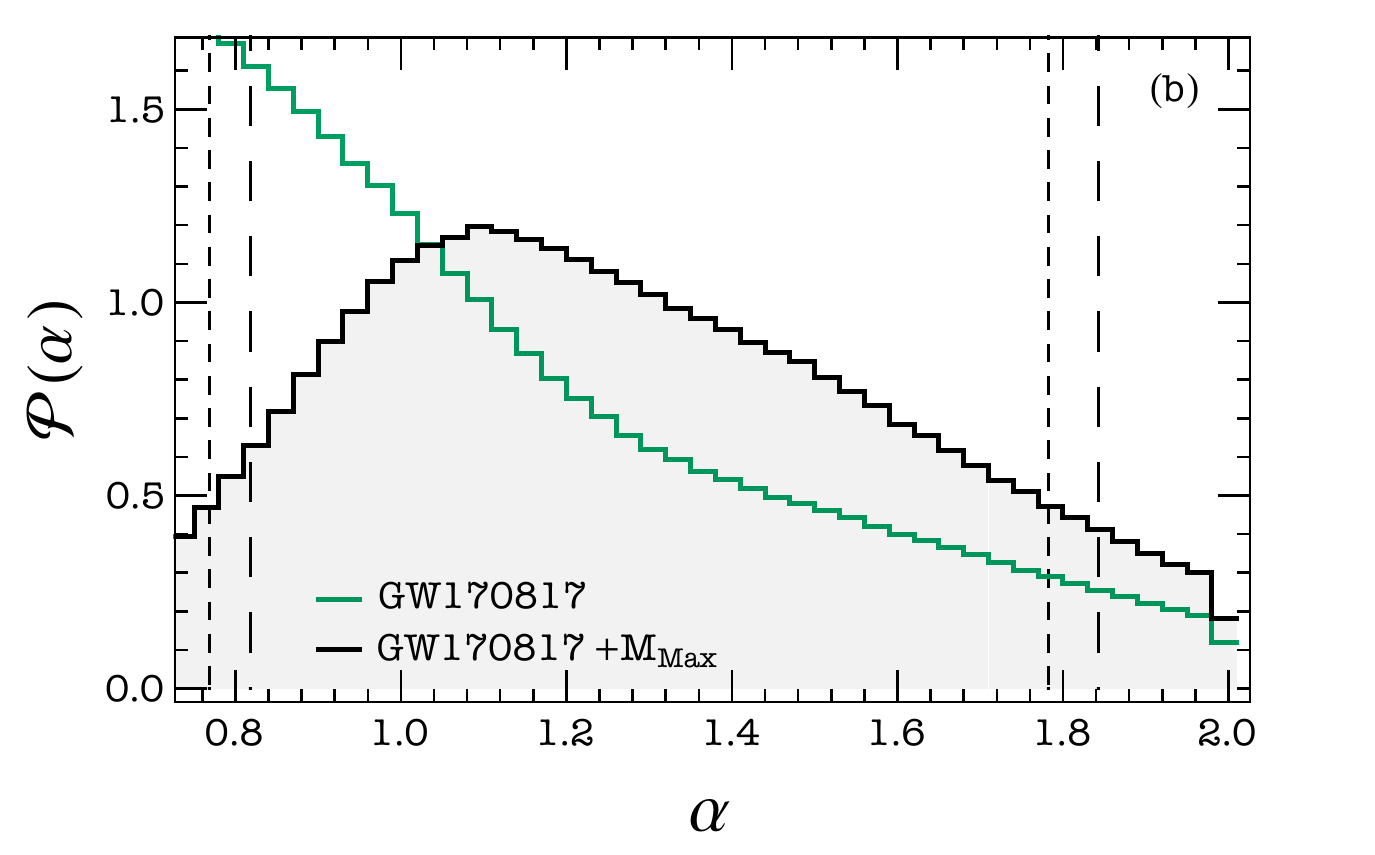}
\includegraphics[width=0.5\textwidth]{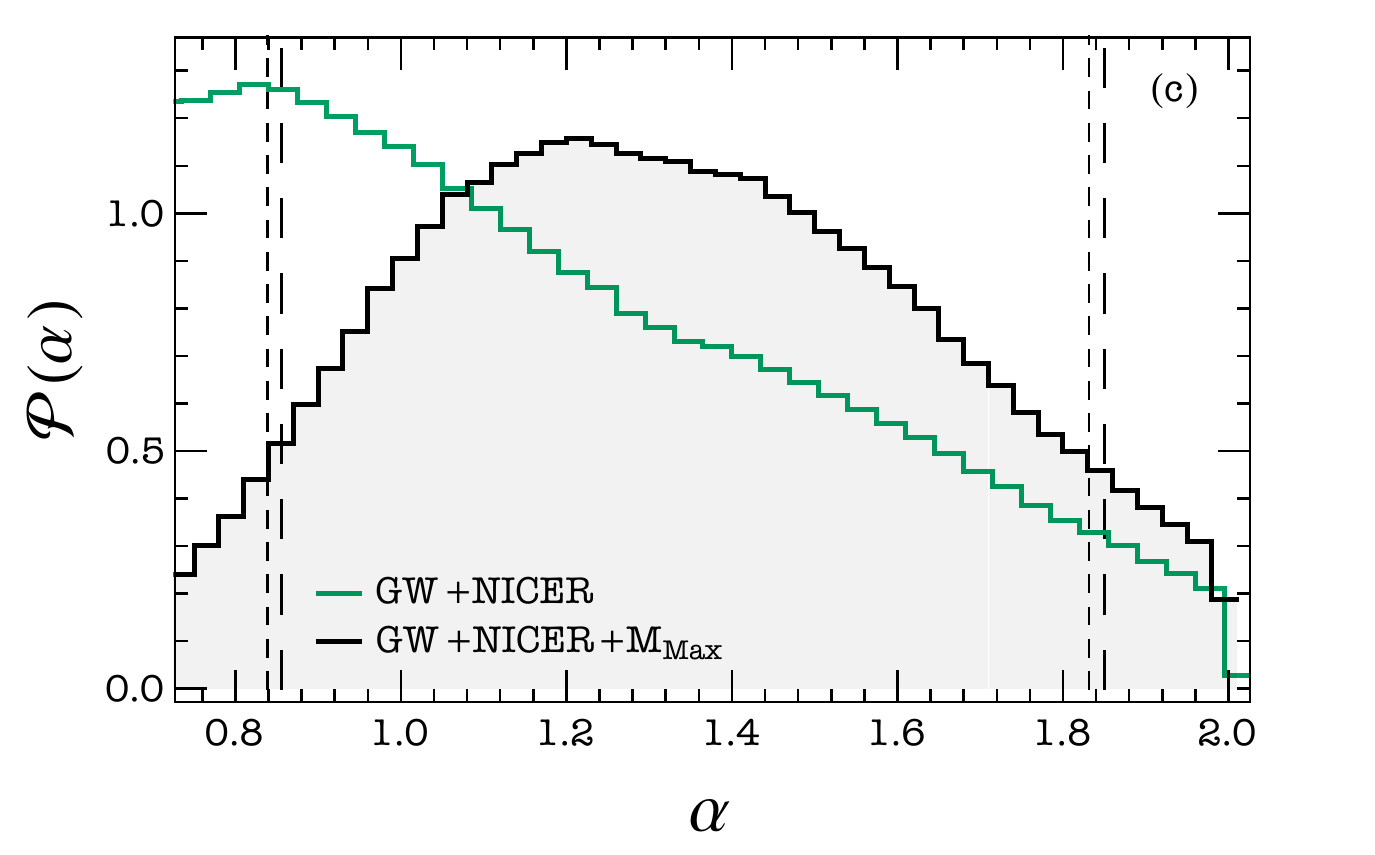}
\caption{Posterior probability distribution of the strength 
of the repulsive three-body potential,  $\alpha$, inferred 
using the mass radius constraints obtained by NICER for 
the millisecond pulsar PSR J0030+0451 (a), the GW observation 
of the binary system GW170817 (b), and combining the two 
datasets (c). Filled and empty histograms correspond to results obtained 
including and neglecting the bound on the maximum mass  
imposed by PSR J0740+6620. Long- and short-dashed vertical 
lines identify 90\% symmetric and highest posterior 
density intervals, respectively.}
\label{fig:2BGpost} 
\end{figure}

We first consider the constraints inferred from either the 
EM or the GW observation alone. Panels (a) and (b) of 
Fig.~\ref{fig:2BGpost} show the marginalized probability 
distribution of $\alpha$ inferred for the two datasets, 
respectively, with and without taking into account the 
maximum mass bound. Interestingly, the GW data alone do 
not seem to provide any reliable constraint on $\alpha$, 
with the marginal posterior being sharply peaked around 
the wall at $\alpha=0.7$ set by the prior. On the other 
hand, inference from the mass-radius measurements of the 
millisecond pulsar observed by NICER yields a tighter 
distribution featuring a peak far from the reference
value $\alpha=1$, and leaves non negligible support for 
the region outside the prior. The inclusion of the bound 
provided by PSR J0740+6620 has a strong effect on both 
observations. This is particular relevant in the case of 
GW170817, in which ${\cal P}(\alpha)$ acquires a well 
defined shape between $\alpha=0.8$ and $\alpha=1.8$. We 
can identify a 90\% confidence interval, which gives 
around the median $\alpha_\tn{GW}=1.25^{+0.48}_{-0.53}$. 
For the EM observation, inclusion of the maximum mass 
bound sharpens the posterior and leads to the value
$\alpha_\tn{EM}=1.52^{+0.43}_{-0.47}$ at 90\% probability. 
Note that NICER points to larger values of $\alpha$, 
which is consistent with the fact that, compared to 
GW170817, the EM measurements seem to favour a stiffer 
EOS.

The multimessenger scenario, illustrated in panel (c), 
suggests that, when no additional data is taken into 
account, the joint NICER\textendash GW inference is 
dominated by the latter.  Including the maximum mass 
constraint results again in a major change, and leads 
to $\alpha_\tn{GW+EM}=1.32^{+0.48}_{-0.51}$ at 90\% 
confidence level. Note that all panels of Fig.~\ref{fig:2BGpost} 
show that the requirement that the EOS support the mass  
of PSR J0740+6620 makes $P(\alpha)$ to have essentially 
zero support for $\alpha\lesssim 0.8$.

\begin{figure}[!htbp]
\centering
\includegraphics[width=0.5\textwidth]{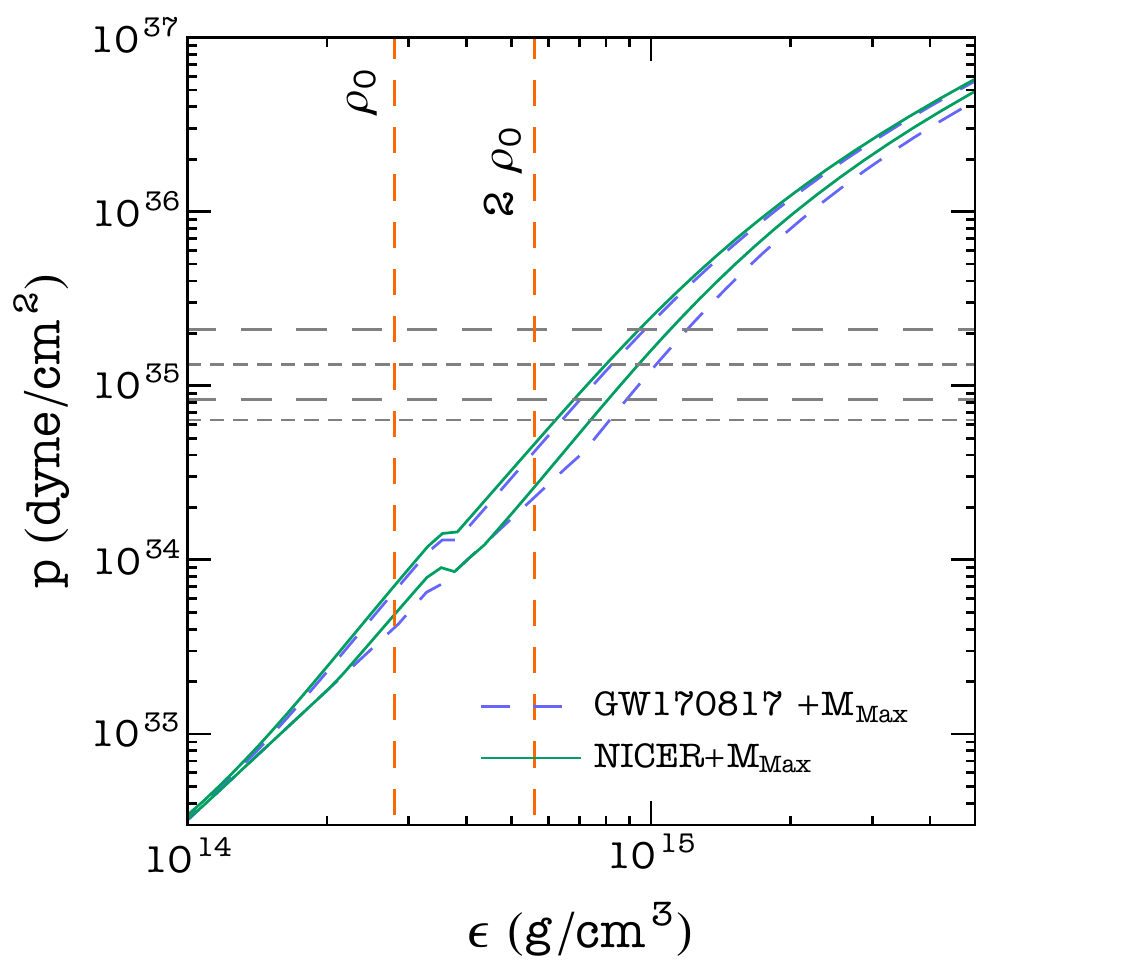}
\caption{Range of pressure-energy density
relations based on the 90\% confidence interval of $\alpha$ 
inferred using GW170817 and PSR J0030+0451, and including the 
maximum mass constraint. Horizontal lines identify the 
90\% high density posterior intervals for the central 
pressures derived for the heavier (dashed) and lighter 
(dotted) NS of GW170817.\label{fig:4BGpost} }
\end{figure}

For the sake of completeness, in Fig.~\ref{fig:4BGpost} 
we also show the range of variation of the the pressure-energy 
density relations obtained using the 90\% high posterior density 
intervals of $\alpha$ resulting from our analysis. Considering that 
the central energy density of the heavier (lighter) NS of GW170817 
obtained from our calculations ranges between $6.2\times10^{14}$g/cm$^3$ 
and $1.2\times10^{15}$g/cm$^3$ ($6.7\times10^{14}$g/cm$^3$ and 
$1.4\times10^{15}$g/cm$^3$) for $0.7 \leq \alpha \leq 2.0$, it
appears that  the EOSs displayed in the figure are largely 
compatible with the empirical bounds 
represented by the horizontal lines. Note that the plateau extending 
between $\sim 1.1 \varrho_0$ and  $\sim 1.5 \varrho_0$ is associated 
with the transition to a phase featuring neutral pion condensation, 
discussed in Ref.~\cite{APR}.

\section{Summary and outlook}
\label{summary}
The quantity and quality of data provided by the new 
observatories operating in the GW and EM band have allowed 
to study the properties of neutron stars evolving in 
different astrophysical environments. These observations 
have set multiple and complementary constraints on the 
NS structure, which have the potential to shed new light 
on different aspects of the microphysics of dense nuclear matter.

We have investigated the constraints that the recent 
observations of GW170817 and the NICER pulsar can impose 
on the potential describing NNN interactions, which are 
presently unconstrained at densities larger than $\sim\varrho_0$. 
Using a Bayesian approach, we have explored both single 
and multimessenger constraints, including also the 
bounds on the maximum mass given by the $2.14 \ M_\odot$ 
pulsar PSR J0740+6620. For the cases analysed, the results 
suggest that constraints on the strength $\alpha$ of  
repulsive NNN interactions are still dominated by the 
requirement that EOS must support the most massive NS 
observed. In this regard, the discovery of NSs more 
massive than PSR J0740+6620 would be potentially able to 
rule out a large part of the parameter space we sampled, 
leading to higher values of $\alpha$, i.e. favoring stiff 
nuclear matter.
The probability distributions inferred for $\alpha$ are 
compatible with the value $\alpha=1$, corresponding to  
the APR EOS~\cite{APR}, providing the baseline for our analysis. 
However, it is interesting to note that the sampled values 
of $\alpha$ show large support for $\alpha>1$, which correspond 
to more repulsive NNN forces and stiffer EOSs.

The work described in this article should be seen as a first step towards  
more systematic studies, aimed at obtaining information on
microscopic nuclear dynamics from the samples of 
observed masses, radii and tidal deformabilities. 

In principle, chiral effective field theory (EFT) provides a 
scheme allowing to derive NN and NNN potentials in a fully consistent
fashion. However, being obtained from a low-momentum expansion, 
chiral potentials cannot be used to obtain the EOS of nuclear matter 
at densities $\varrho \gg \varrho_0$~\cite{Benhar_EFT}. 
In particular the breakdown scale of EFT is expected to be between 
1-2$\varrho_0$. On this respect the authors of ~\cite{Essik:2021} 
explored the possibility of using multimessenger astronomy to test 
the predictions of EFT for the EOS only up to $2\varrho_0$. 

Further applications of the methodology underlying our approach can be 
pursued following different directions, and in particular 
considering observations of binary NS by LIGO/Virgo and the Japanese 
detector KAGRA \cite{Somiya:2011np,Akutsu:2018axf} at design 
sensitivity, as well as by third generation interferometers, such as 
the Einstein Telescope \cite{Maggiore:2019uih,Sathyaprakash:2019yqt,Punturo:2010zz}. 
In this context, it should be kept in mind that 
the accuracy of the bounds inferred from GW170817 are expected to improve 
by at least a factor $\sim 3$ with LIGO/Virgo at full capacity, and by 
more than one order of magnitude with the planned Einstein 
Telescope \cite{GuerraChaves:2019foa}. The large rate of detections provided 
by these instruments will allow to stack multiple observations and to 
study nucleon interactions from constraints 
on sources within a wider mass spectrum. These data,  
supplemented by electromagnetic measurements that will also benefit from 
longer observation times \cite{Riley:2019yda}, will narrow the 
bounds on $\alpha$, with the errors expected to scale as the number 
of events increases.
Finally, the quality and the quantity of data obtained from the 
new facilities will allow to test more sophisticate models and to 
explore, for example, the sensitivity of GW and EM signals to an 
explicit density-dependence of the strength of the 
three body potential.

\begin{acknowledgments}
The authors are indebted to Valeria Ferrari, Margherita 
Fasano, and Alessandro Lovato for a critical reading of this 
manuscript. This work has been supported by INFN under grant 
TEONGRAV. 
A.M. acknowledges support from the Amaldi Research Center,  
funded by the MIUR program ``Dipartimento di Eccellenza'',  
CUP: B81I18001170001.
\end{acknowledgments}

\bibliographystyle{apsrev4-2}
\bibliography{Ref}

\end{document}